\documentclass[sigconf, anonymous=false,nonacm]{acmart}

\AtBeginDocument{%
  }

\setcopyright{acmlicensed}
\copyrightyear{2018}
\acmYear{2018}
\acmDOI{XXXXXXX.XXXXXXX}
\acmConference[Conference acronym 'XX]{Make sure to enter the correct
  conference title from your rights confirmation email}{June 03--05,
  2018}{Woodstock, NY}
\acmISBN{978-1-4503-XXXX-X/2018/06}

\usepackage[utf8]{inputenc} % allow utf-8 input
\usepackage[T1]{fontenc}    % use 8-bit T1 fonts
\usepackage{hyperref}       % hyperlinks
\usepackage{url}            % simple URL typesetting
\usepackage{booktabs}       % professional-quality tables
\usepackage{amsfonts}       % blackboard math symbols
\usepackage{enumitem}       % customize enumerate environment
\usepackage{nicefrac}       % compact symbols for 1/2, etc.
\usepackage{microtype}      % microtypography
\usepackage{xcolor}         % colors
\usepackage{graphicx} % Required for inserting images
\usepackage{multirow}
\usepackage{hyperref}
\usepackage{amsmath}
\usepackage{geometry}
\usepackage{xcolor} % Required for cell coloring
\usepackage{colortbl} % Required for \cellcolor
\usepackage{booktabs} % For better table formatting
\usepackage{todonotes}
\usepackage{subcaption}
\usepackage{multicol}
\usepackage{balance}

\newcommand{\smallhtwoo}{\ensuremath{8-\texttt{H}_2\texttt{O}}}
\newcommand{\bightwoo}{\ensuremath{10-\texttt{H}_2\texttt{O}}}

\newcommand{\lih}{\ensuremath{6-\texttt{LiH}}}
\newcommand{\behtwo}{\ensuremath{6-\texttt{BEH}_2}}

\definecolor{qaserc}{RGB}{255, 177, 140}
\definecolor{nqaserc}{RGB}{117, 212, 184}

\begin{document}

\title{QASER: Breaking the Depth vs. Accuracy Trade-Off\\for Quantum Architecture Search}

\author{Ioana Moflic}
\email{ioana.moflic@aalto.fi}
\affiliation{%
  \institution{Aalto University}
  \city{Espoo}
  \country{Finland}
}

\author{Alexandru Paler}
\email{alexandru.paler@aalto.fi}
\affiliation{%
  \institution{Aalto University}
  \city{Espoo}
  \country{Finland}
}

\author{Akash Kundu}
\email{akash.kundu@helsinki.fi}
\affiliation{%
  \institution{University of Helsinki}
  \city{Helsinki}
  \country{Finland}
}

\renewcommand{\shortauthors}{Moflic et al.}

\begin{abstract}
Quantum computing faces a key challenge: balancing the need for low circuit depth (crucial for fault tolerance) with the high accuracy required for complex computations like quantum chemistry and error correction, which typically require deeper circuits. We overcome this trade-off by introducing a novel reinforcement learning approach featuring engineered reward functions, called \textbf{QASER}, that take into account seemingly contradictory optimization goals. This reward enables the compilation of circuits with lower depth and higher accuracy, significantly outperforming state-of-the-art techniques. Benchmarks on quantum chemistry state preparation circuits demonstrate stable compilations. We achieve up to 50\% improved accuracy, while reducing 2-qubit gate counts and depths by 20\%. This advancement enables more efficient and reliable quantum compilation. Source code is available at \href{here}{[removed for review]}.
\end{abstract}

% \begin{CCSXML}
% <ccs2012>
%    <concept>
%        <concept_id>10010583.10010786.10010813.10011726</concept_id>
%        <concept_desc>Hardware~Quantum computation</concept_desc>
%        <concept_significance>500</concept_significance>
%        </concept>
%    <concept>
%        <concept_id>10010583.10010682.10010689</concept_id>
%        <concept_desc>Hardware~Hardware description languages and compilation</concept_desc>
%        <concept_significance>500</concept_significance>
%        </concept>
%    <concept>
%        <concept_id>10003752.10010070.10010071.10010261</concept_id>
%        <concept_desc>Theory of computation~Reinforcement learning</concept_desc>
%        <concept_significance>500</concept_significance>
%        </concept>
%  </ccs2012>
% \end{CCSXML}

% \ccsdesc[500]{Hardware~Quantum computation}
% \ccsdesc[500]{Hardware~Hardware description languages and compilation}
% \ccsdesc[500]{Theory of computation~Reinforcement learning}


\maketitle

\section{Introduction}

Quantum computing holds the promise of solving complex problems that are intractable for classical computers~\cite{kim2023evidence}. A key challenge in realizing this promise lies in the design and optimization of quantum circuits. Quantum architecture search~\cite{zhang2022differentiable} (QAS) has emerged as a promising field to automate the design of efficient quantum circuits. For example, QAS can automatically discover the gate sequence needed to prepare a maximally entangled state~\cite{du2022quantum, fosel2021quantum}, or find the arrangement of rotation gates needed for a circuit to approximate the ground state of a Hamiltonian~\cite{ostaszewski2021reinforcement}. Finding the ground state of such an electronic-structure Hamiltonian is know to fall under QMA-complete problems~\cite{o2021electronic}. SAT-based methods~\cite{jakobsen2025depth}, although guaranteed to offer the optimum circuit, suffer from severe scalability limitations—becoming computationally infeasible for larger qubit counts or deep circuits. As such, reinforcement learning (RL) has emerged as a promising approach for navigating the vast and complex space of circuit architectures~\cite{quetschlich2023compiler, weiden2023improving, fernandes2025enhancing}. RL techniques have similarly found success in classical design~\cite{huang2021machine,markov2024reevaluating} and error-corrected quantum hardware calibration~\cite{sivak2025reinforcement}.

In the RL-based QAS framework, an agent sequentially constructs circuits by selecting gates and their placements, using feedback from the cost function as a reward signal to guide policy updates. However, one of the persistent challenges in RL is reward engineering~\cite{gupta2022unpacking, dewey2014reinforcement}. Designing an appropriate reward function that effectively captures circuit quality, resource efficiency, and physical constraints is nontrivial. Poorly shaped rewards can lead to unstable training or suboptimal exploration of the search space~\cite{devidze2022exploration}.

The area of reward engineering for QAS remains relatively niche and less explored. This gap is particularly consequential, as circuits that perform well theoretically may still be highly susceptible to errors when implemented on physical hardware. This paper contributes to that direction by introducing a novel reward engineering technique that mitigates key limitations in existing approaches and moves toward scalable, fault-tolerant architectures for the post-NISQ era~\cite{zimboras2025myths, preskill2025beyond, eisert2025mind}.

\begin{figure}[t!]
    \centering
    \includegraphics[width=0.8\linewidth]{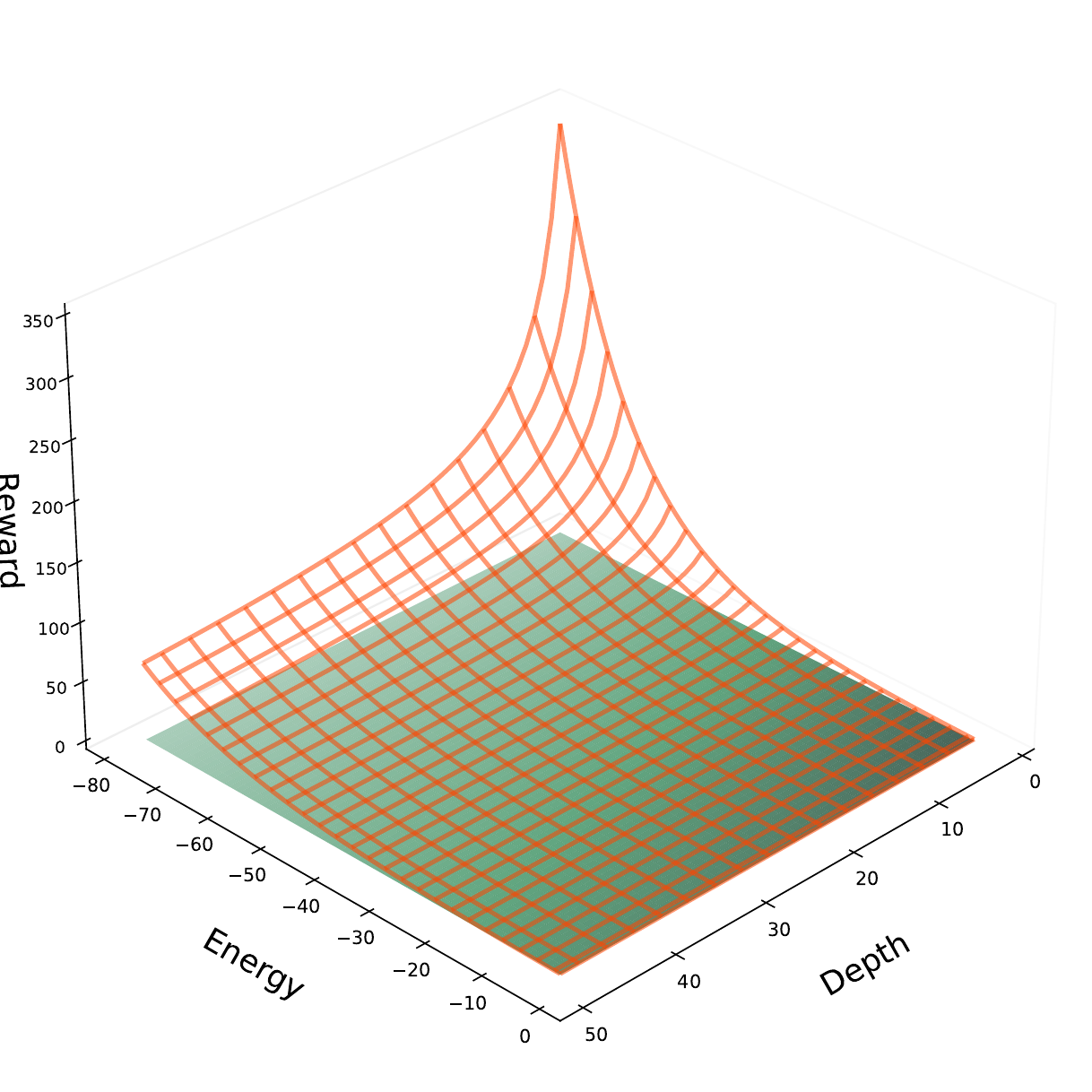}
    \caption{Our QASER exponential reward function is denser than those commonly used in the RL literature for QAS (e.g. the green surface). QASER (the orange meshed surface, e.g. Eq.~\ref{eq:exp_reward_orig}) captures multiple costs, offering a carefully tailored reward signal to the RL agent. The exponential nature of QASER ensures that the best reward signals are for circuits which exhibit simultaneous low depth and low energy.}
    \label{fig:rewards}
\end{figure}

\subsection{Quantum architecture search (QAS)}

A wide range of QAS strategies have been developed to optimize parameterized quantum circuits (PQCs), each leveraging distinct algorithmic paradigms~\cite{martyniuk2024quantum, lu2023qas}. Adaptive methods, such as~\cite{huang2024adaptive} are proposed in the selection of pool operators in ground state estimation, while Monte Carlo sampling approaches have enabled the discovery of PQCs for tasks such as quantum Fourier transform and Max-Cut~\cite{zhang2022differentiable}. Supernet-based weight sharing has been explored in quantum chemistry~\cite{du2022quantum}, although these methods are often sensitive to noise. Evolutionary algorithms~\cite{ding2022evolutionary}, genetic algorithms~\cite{xie2025deqompile}, and representation learning~\cite{sun2024quantum}, have also shown promise in circuit construction. Differentiable search techniques, including~\cite{zhang2022differentiable}, enable gradient-based optimization of circuit architectures, and training-free QAS methods~\cite{he2024training} based on path and expressivity metrics have recently been introduced. Recent advances have significantly broadened the landscape of quantum circuit optimization; however, scalability remains a persistent challenge, with most existing approaches only addressing problems involving a limited number of qubits. Authors of~\cite{zhu2025scalable} demonstrated QAS for up to 12-qubits by leveraging matrix product states and detailed landscape analysis. In contrast, standard approaches such as TF-QAS~\cite{he2024training} and DQAS~\cite{zhang2022differentiable, wu2023quantumdarts} rapidly encounter memory bottlenecks, rendering both the search process and subsequent circuit evaluation infeasible for systems exceeding this scale.

\subsection{Reinforcement learning (RL) for QAS}

The use of RL for QAS has become a heavily researched topic in recent years. Typically, it focuses on minimizing a specific cost function, such as circuit performance, and/or metrics like circuit gate count or depth~\cite{fosel2021quantum}. In this framework, agents are guided by tailored reward functions to identify effective quantum gate sequences. RL algorithms such as double-deep Q-networks~\cite{van2016deep} (DDQN) with $\epsilon$-greedy policies, have been used for ground state estimation~\cite{ostaszewski2021reinforcement}, while actor-critic and proximal policy optimization algorithms have facilitated the generation of multi-qubit entangled states~\cite{kuo2021quantum}. RL has further contributed to hardware-aware circuit design and efficiency improvements~\cite{fosel2021quantum}, with recent works incorporating curriculum learning and pruning for QAS~\cite{patel2024curriculum}. 

Recent RL-based QAS scaled up to 20-qubits by leveraging tensor network representations which efficiently initialize the search and constrain the optimization landscape~\cite{kundu2025tensorrlqas}. While this approach address some scalability limitations imposed by quantum simulators, the primary computational bottleneck lies in the repeated simulator queries, especially the iterative extraction of the state vector at every RL-step. Methods such as GadgetRL~\cite{kundu2024reinforcement, olle2025scaling} tried to address this issue by focusing on sub-circuits (\emph{gadgets}) as expanded actions, but the vast architectural space is still a bottleneck.

RL-QAS has further been used for optimizing fault-tolerant quantum circuits~\cite{zen2025quantum}. Although the simulators in this setting are efficient, the size of the architectural search space continues to pose a significant bottleneck. The structure of a quantum circuit significantly influences its robustness: under realistic noise models, entangling gates such as the \texttt{CNOT} gate can propagate 1-qubit errors throughout the circuit, thereby degrading its fault-tolerance. In RL-QAS specifically, fault-tolerance considerations are still underexplored and are rarely embedded in the reward design. Moreover, across RL-QAS the fundamental trade-off between solution accuracy and circuit gate count continues to pose a significant challenge, representing an unresolved barrier for scalability and performance.

\subsection{Reward engineering for RL-based QAS}

Reward engineering refers to the systematic design, shaping, and refinement of reward functions to ensure that an agent's behavior aligns with the intended objectives of the task. Reward engineering plays a crucial role in the success of the RL algorithm. Effective reward engineering improves the stability and sample efficiency of learning, particularly in environments where rewards are sparse, ambiguous, or difficult to specify directly. A major open challenge~\cite{altmann2024challenges} is to develop a smoother, more informative reward metrics to facilitate RL-agent exploration in complex circuit spaces. Addressing this is vital for advancing RL-based QAS.

In the context of QAS, the typical rewards were very simple. In~\cite{kuo2021quantum}, the authors use a sparse reward in which the agent receives good feedback only when the fidelity of the circuits passes a certain threshold. The reward function of Patel et al. \cite{patel2024curriculum} takes the estimated energy of the circuit as a parameter and incorporates two extreme reward values for two different stopping conditions (exceeding error threshold and reaching the maximum number of actions). The work in~\cite{fosel2021quantum} employs a reward function which is a linear combination of gate count and depth. In~\cite{Nägele_2024}, the reward function is a step-wise difference between the number of nodes in the ZX-graph to optimize.

Authors of~\cite{kuo2021quantum} present a sparse reward in which the agent receives positive feedback only when the state of the circuit proposed by RL-agent matches a multi-qubit maximally entangled state. Ref.~\cite{kundu2024enhancing} introduce a logarithmic reward function for quantum state diagonalization where if the agent fully diagonalizes the state it receives a positive high response and the intermediate rewards are given by a logarithmic function. More advanced rewards have only recently been introduced. In~\cite{veviurkotothemax}, authors use a novel reward function design in reinforcement learning, where the task of the agent is to optimize the maximum reward, not the cumulative one. This approach shares some similarities with our method in terms of relying on maximum values of costs encountered during the learning process. However, our approach differs in how these values are incorporated into the reward and the specific application to the simultaneous optimization of energy and depth in quantum chemistry circuits.

\subsection{Contribution}

We introduce QASER, a framework driven by an exponential reward function for reinforcement learning-based QAS (Fig.~\ref{fig:rewards}). Unlike previous RL-QAS~\cite{ostaszewski2021reinforcement, patel2024curriculum}, which primarily rely on sparse or linear rewards and often sacrifice efficiency for accuracy, QASER leverages a max-tracking mechanism to drive steady reductions in all targeted costs (such as circuit depth, number of 2-qubit gates and accuracy).

On quantum chemistry benchmarks, QASER achieves up to 20\% fewer 2-qubit gates and reduced circuit depth with better accuracy compared to state-of-the-art QAS. The improvement stems directly from advanced reward engineering and demonstrates that a principled, multi-objective reward is key to pushing RL-QAS beyond the limitations of previous approaches.

The rest of the sections are arranged as follows. We find a trade-off between the commonly encountered costs in QAS (Section~\ref{sec:costs}), such as the ones from PQC (e.g. estimated energy) and the ones from quantum circuit optimization (e.g. gate count and depth). We demonstrate the effectiveness of the reward function in both noisy (Section~\ref{sec:noisy}) and noiseless scenarios (Section~\ref{sec:noiseless}). Moreover, we investigate the effectiveness of one of the variants of the exponential reward function in TensorRL-QAS (Section~\ref{sec:warm}), which is well suited for the specific framework.

\section{Exponential Rewards}
\label{sec:methods}

Baseline reinforcement learning for quantum circuit optimization~\cite{ostaszewski2021reinforcement, kundu2024reinforcement, patel2024curriculum} uses an objective-oriented reward function to accelerate training and discover more compact quantum circuits in noisy scenarios. Typically, it focuses on energy minimization and has rewards defined as:
\begin{equation}
R =
\begin{cases}
r, & \text{if } E_t < \zeta, \\
-r & \text { if } t \geq D_\text{max} \text { and } E_t \geq \zeta, \\
\mathcal{E}, & \text{otherwise}.
\end{cases}
\label{case:error_rwd_CRLQAS_vanilla}
\end{equation}
where $r=5$, $\mathcal{E} = \text{max}\left(\frac{E_{t-1} - E_t}{\lvert E_{t-1} - E_\text{min} \rvert}, -1\right)$, $E_t$ is the value of the cost function at time step $t$, $\zeta$ is predefined threshold\footnote{In our case the chemical accuracy $1.6\times10^{-3}$ Hartree} and $D_\text{max}$ is the maximum number of allowed steps (i.e., maximum circuit depth) per episode.

While this reward structure achieves $\zeta$ by incentivizing early convergence and penalizing excessive depth, it has several limitations. First, it lacks explicit incentives for circuit efficiency, creating a suboptimal exploration-exploitation tradeoff. Second, the agent may overfit to energy reduction at the expense of circuit complexity, as intermediate rewards for energy improvements do not distinguish between necessary and redundant gates. This can lead to \emph{reward hacking}~\cite{memarian2021self}, where agents exploit small energy fluctuations by adding unnecessary gates, inflating circuit depth. Third, the abrupt penalty at $D_\text{max}$ may discourage late-stage exploration, whereas softer penalty schedules could better balance depth constraints with final-stage optimization.

\begin{table*}[t!]
\small
\setlength{\tabcolsep}{10pt}

\begin{tabular}{@{}llccc lccc@{}}
\toprule

& & \multicolumn{3}{c}{\textbf{QASER}} 
& & \multicolumn{3}{c}{\textbf{CRLQAS}}  \\
& & $\lih$ & $\smallhtwoo$ & $\bightwoo$ 
& & $\lih$ & $\smallhtwoo$ & $\bightwoo$ \\
\midrule

% ---- Avg Error ----
Avg.\ Error &
& \cellcolor{qaserc!80}\textbf{$6.5\times 10^{-5}$}
& \cellcolor{qaserc!80}\textbf{$4.3\times 10^{-4}$}
& \cellcolor{qaserc!80}\textbf{$3.4\times 10^{-4}$}
& Avg.\ Error
& $8.39\times 10^{-5}$ 
& $8.77\times 10^{-4}$ 
& $3.53\times 10^{-4}$ \\

% ------------------ CNOT --------------------
Max.\ CNOT &
& 54 
& \cellcolor{qaserc!80}\textbf{133} 
& 224
& Max.\ CNOT
& \cellcolor{nqaserc!80}\textbf{52} 
& 193 
& \cellcolor{nqaserc!80}\textbf{187} \\

Avg.\ CNOT &
& \cellcolor{qaserc!80}\textbf{28.4}
& \cellcolor{qaserc!80}\textbf{81.9}
& \cellcolor{qaserc!80}\textbf{103.1}
& Avg.\ CNOT
& 40.2 
& 99.1
& 112.5 \\

Min.\ CNOT &
& \cellcolor{qaserc!80}\textbf{13}
& 35
& \cellcolor{qaserc!80}\textbf{34}
& Min.\ CNOT
& 20
& \cellcolor{nqaserc!80}\textbf{27}
& 44 \\

% ------------------ Depth --------------------
Max.\ Depth &
& 63
& \cellcolor{qaserc!80}\textbf{129}
& 178
& Max.\ Depth
& \cellcolor{nqaserc!80}\textbf{54}
& 193
& \cellcolor{nqaserc!80}\textbf{143} \\

Avg.\ Depth &
& \cellcolor{qaserc!80}\textbf{38.2}
& \cellcolor{qaserc!80}\textbf{78.6}
& \cellcolor{qaserc!80}\textbf{81.3}
& Avg.\ Depth
& 39.5 
& 91.1
& 86.6 \\

Min.\ Depth &
& \cellcolor{qaserc!80}\textbf{18}
& 42
& 37
& Min.\ Depth
& 20
& \cellcolor{nqaserc!80}\textbf{22}
& 37 \\

\bottomrule
\end{tabular}

\vspace{5mm}
\caption{
\textbf{Comparison of QASER and CRLQAS on $\lih$, $\smallhtwoo$, and $\bightwoo$.}  
Cells highlighted in \textcolor{qaserc!80}{orange} indicate best-performing values for QASER. Cells highlighted in \textcolor{nqaserc!80}{green} indicate best-performing values for CRLQAS in the same benchmarks.
}
\label{tab:noiseless_comparison}
\end{table*}

\subsection{Quantum circuit costs}
\label{sec:costs}

We aim to encode the following costs into our reward function: \emph{energy}, \emph{depth}, and \emph{gate cost}. By incorporating these additional cost terms, our approach provides a more nuanced and practical reward signal, guiding the RL-agent to generate quantum circuits that are not only theoretically optimal, but also robust and efficient in real-world, noisy quantum environments.

\textbf{Observation:} 2-qubit gates have lower gate fidelity than 1-qubit gates. For the purpose of this work, we consider $n=2$ classes of gates, however $n$ can be arbitrarily large depending on the noise model or quantum architecture. We model this property by assigning different gate cost weights to the $n = 2$ classes of gates: 1) 1-qubit (\texttt{RX}, \texttt{RY}, \texttt{RZ}) and 2) 2-qubit (\texttt{CNOT}) gates. The weighted gate cost is motivated by the need to both reduce the error rate of the circuit, and to prevent the RL-agent from adding redundant entangling gate sequences in the circuit. 

\textbf{Definition:} The \textit{gate cost} is an estimate of the cost $C$ of the circuit $s_t$ in terms of the gates it is composed of and is computed as a weighted average of the gates in the circuit:
\[
C(s_t) = \frac{\sum_{i=1}^{n}{w_i x_i}}{\sum_{i=1}^{n}{w_i}}.
\]
where $w_i$ is the assigned weight of each of the $n$ gate classes, and $x_i$ is the number of gates in circuit at state $s_t$ that are of class $i$.

\textbf{Observation:} Reducing the gate cost of the circuit can be correlated with a better depth, therefore these two values can sometimes be simultaneously minimized in the reward function.

\textbf{Definition: } 
The \textit{depth} is an estimate of the depth of the circuit $s_t$ in terms of the gates it is composed of. It is computed as a weighted average, similar to the gate cost, since in fault-tolerant scenarios different gates in $s_t$ may contribute unequally to the overall depth. Herein, without loss of generality, we consider the simple case of a single class of gates having the same depth weight. 

In QAS for Hamiltonian ground state preparation, the RL environment is defined by parameterized quantum circuits (PQCs) which represent molecular ground states. At each timestep~$t$, we evaluate the estimated energy~$E_{s_t}$ of the circuit state~$s_t$, aiming to approach the ground state energy~$E_{\min}$ of the molecule.

\textbf{Observation:} For other QAS tasks, such as compilation of fault-tolerant circuits~\cite{preskill2025beyond,zen2025quantum}, the energy can be replaced with other values, such as the fidelity of the output state.

\textbf{Definition: } The \emph{energy} refers to the ground state energy of the molecular Hamiltonian that is being compiled as a circuit.

\subsection{QASER: Combining costs into a reward}

Our approach, QASER, to designing the reward function is inspired by the one proposed in~\cite{veviurkotothemax}. We will use the shorthand notation $E(s_t), D(s_t)$ and $C(s_t)$ to refer to the energy, depth, and gate cost of the circuit $s_t$. We introduce two auxiliary variables, $\mathbb{M}_{D, t}$ and $\mathbb{M}_{C, t}$, which track the maximum depth and gate cost encountered along a learning trajectory up to time step $t$.

\textbf{Definition:} The QASER reward function at time step \( t \) is:

\begin{equation}
R(s_t)_{QASER} = \left(\frac{\mathbb{M}_{D, t}}{D(s_t) + 1} + \frac{\mathbb{M}_{C, t}}{C(s_t) + 1}\right)^{\frac{E(s_t)}{E_{min}}},
\label{eq:exp_reward_orig}
\end{equation}

where each auxiliary variable \( \mathbb{M}_{f, t} \) is given by:
\begin{equation}
\mathbb{M}_{f, t} = \max \left\{ f(s_k) \,\middle|\, 0 \leq k < t \right\}, \quad \text{for } f \in \{\text{C}, \text{D}\}.
\label{eq:max_tracking}
\end{equation}

The definition of the function in Eq.~\ref{eq:exp_reward_orig} ensures that the reward depends not only on the current state but also on the historical worst-case values along the trajectory up to the current time step $t$. 
In summary, our reward function:
\begin{itemize}
    \item \textit{incorporates multiple costs} which reflect different properties of the same PQC;
    \item \textit{uses a max-based tracking mechanism}, in Eq.~\ref{eq:max_tracking}, which propagates the worst observed circuit costs through the trajectory via iterative updates of the form \( \mathbb{M}_{f, t} = \max\{ \mathbb{M}_{f, t-1}, f(s_t) \} \); 
    \item \textit{has an exponential structure}, rewarding PQCs that simultaneously exhibit desirable properties such as low estimated energy, low gate count, and shallow depth.
\end{itemize}

\subsection{Bounding the reward}
\label{sec:bound}

The QASER reward increases exponentially when all costs are simultaneously as low as possible compared to the maximum encountered costs. Tracking historical maxima \( \mathbb{M}_{f, t} \) allows the reward to reflect how the agent's performance at time $t$ compares to the most expensive states encountered in previous trajectories, encouraging a consistent improvement. In this exponential structure, each $\mathbb{M}_{f, t} / (f(s_t) + 1)$ term of Eq.~\ref{eq:exp_reward_orig} adjusts the reward relative to the worst-case cost observed up to that point. When $\mathbb{M}_{f, t} / (f(s_t) + 1) \leq 1$, the cost function $f$ has reached a new maximum (reward is decreasing) and when $\mathbb{M}_{f, t} / (f(s_t) + 1) > 1 $, $f$ is in a new minimum (reward is increasing). 

One may wonder why the exponent does not follow the same \textit{max}-based tracking rule from Eq.~\ref{eq:max_tracking}. Part of the reasoning is motivated by the need to avoid infinite reward values (Sec.~\ref{sec:bound}). However, we also have more practical considerations: while we have an estimate of the ground-state energy $E_{min}$ for a given molecule, there is no known estimation of what the minimum gate count and depth of the optimal PQC might be. Without such information, instead of asking ourselves how far from the absolute minimum the current solution is, we revert the question and ask \textit{how well do we perform in state $s_t$ compared to the worst historical $s_{k<t}$}?

The initialization of the values $\mathbb{M}_{f, 0} $ plays a critical role in ensuring that rewards in the early time steps (e.g, $t=1$) are meaningful. 
A realistic and problem-informed initialization improves the reward scaling across the trajectory and results in more stable learning.
\begin{figure*}[t!]
    \centering
    \begin{subfigure}{0.49\textwidth}
        \centering
        \includegraphics[width=1\linewidth]{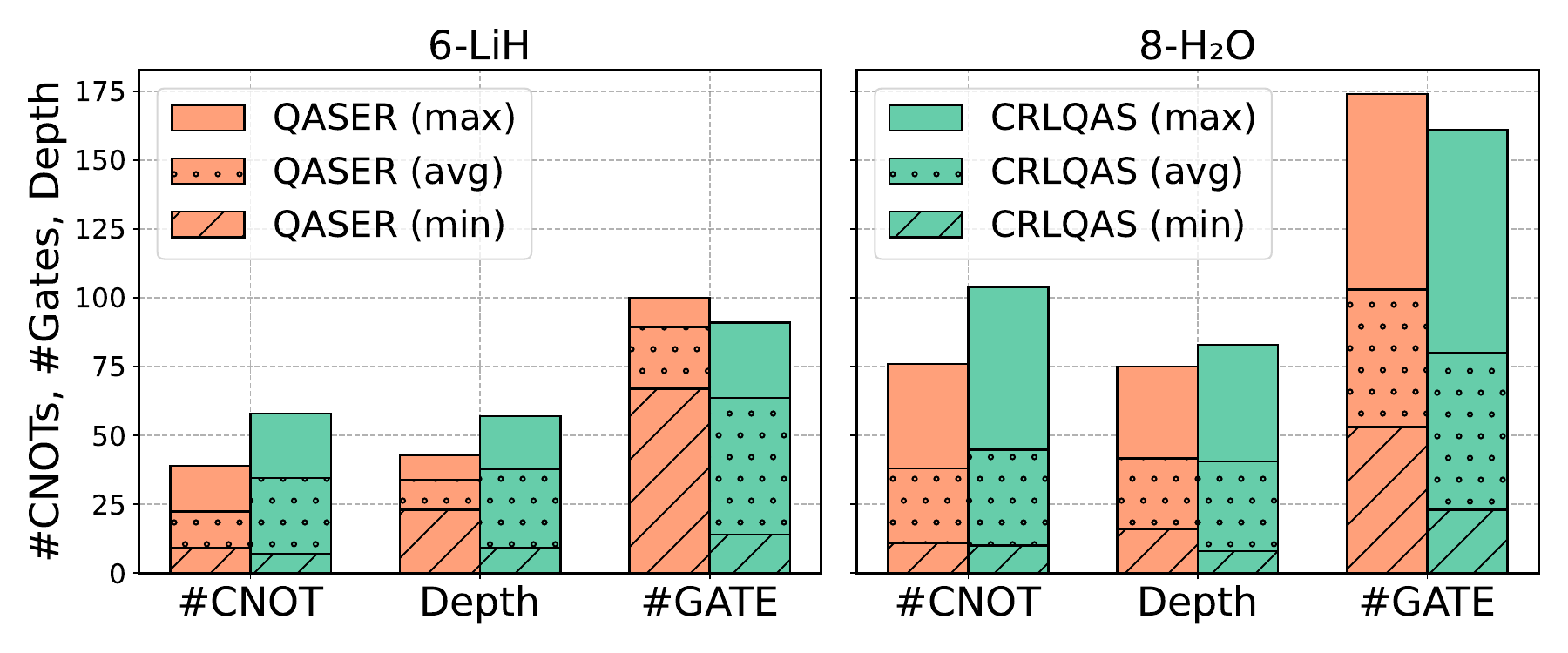}
        \caption{  }
        \label{fig:noisy_simulation}
    \end{subfigure}
    \hfill
    \begin{subfigure}{0.49\textwidth}
        \centering
        \includegraphics[width=1\linewidth]{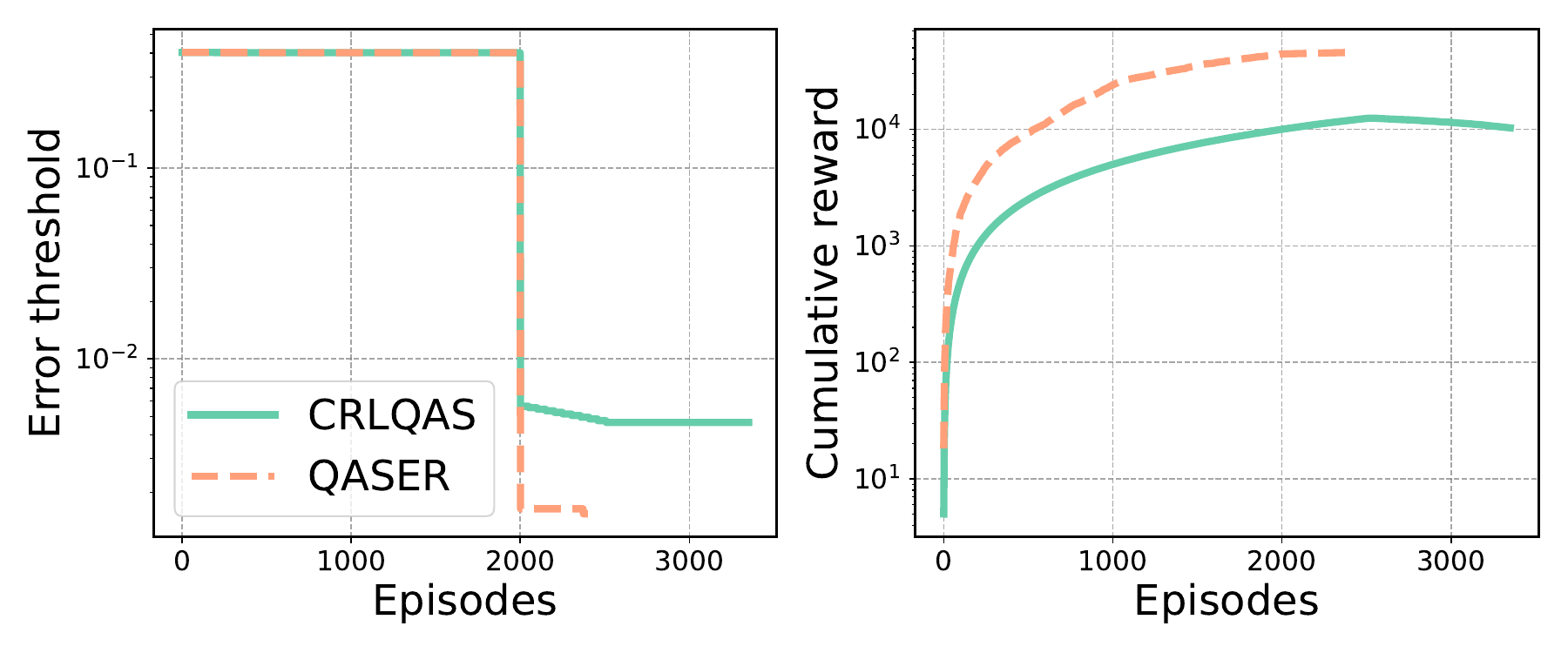}
        \caption{ }
        \label{fig:noisy_simulation_performance}
    \end{subfigure}
    \caption{QASER vs CRLQAS. (a) QASER outperforms CRLQAS in the realistic noisy scenario in finding the ground state of $\lih$, and $\smallhtwoo$ molecules. (b) QASER exhibits accelerated reward accumulation, converging to approximately $10^{4}$ reward units, while CRLQAS demonstrates more gradual learning dynamics, reaching a plateau at approximately $10^3$. QASER shows faster convergence and a better reward signal when compared to CRLQAS in finding the ground state of a $\bightwoo$ molecule.}
    \label{fig:main}
\end{figure*}

\textbf{Lemma:} The QASER reward function (Eq.~\ref{eq:exp_reward_orig}) is bounded (under practically reasonable assumptions).

\textbf{Proof:} We analyze the extreme cases to demonstrate that 
QASER is bounded. Suppose that the initial maximum values \( \mathbb{M}_{f, 0} \) are non-infinite and high enough such that \( \mathbb{M}_{f, 0} \geq f(s_k) \) for all \( 0 \leq k < t \). For a given energy estimate \( E(s_t) \), two boundary cases arise:
\[
\lim_{E(s_t)\to 0} R(s_t)=1,\qquad 
\lim_{E(s_t)\to E_{\min}} R(s_t)=\frac{\mathbb{M}_{D,t}}{D(s_t)+1}+\frac{\mathbb{M}_{C,t}}{C(s_t)+1}.
\]

This implies that the reward function \( R: \mathcal{S} \rightarrow (1, \bar{R}) \) is non-negative and bounded, with the upper limit determined by the historical maximums of the cost functions.

\section{Results}

Herein, we evaluate QASER on quantum chemistry problems ranging from 6- to 10-qubit. Specifically targeting the ground state preparation of $\behtwo$\footnote{Molecules and their corresponding qubit requirements are labeled in the format ``$N$-molecule\_name'', where 
$N$ indicates the number of qubits.}, $\smallhtwoo$ (both in STO-3G) and $\bightwoo$ (in 6-31G), under cold-start, and warm-start QAS framework. The latter is when the RL-agent starts from prior knowledge in order to accelerate training~\cite{kundu2025tensorrlqas}. In Section~\ref{sec:cold}, we show that for cold-start QAS, QASER achieves lower-depth, highly-accurate circuits compared to state-of-the art CRLQAS~\cite{patel2024curriculum}. In Section~\ref{sec:warm}, we show that QASER can significantly accelerate the state-of-the-art warm-start TensorRL-QAS~\cite{kundu2025tensorrlqas}. In both situations, we consider 10 random initializations of the neural network in order to remove stochasticity and to assess the stability of a RL-agent.% To generate the chemical Hamiltonian, we use \texttt{Pennylane}~\cite{bergholm2018pennylane} and \texttt{OpenFermion}~\cite{mcclean2020openfermion}.  

In the following, QASER demonstrates dual benefits (1) it improves simulation accuracy while simultaneously reducing quantum resource requirements. This characteristic is particularly valuable for post-NISQ-era~\cite{preskill2025beyond}, where gate errors accumulate due to circuit size, and not due to high intrinsic gate error rates. (2) The consistency of performance advantages across systems of varying complexity suggests the robustness of our approach to problem scale.

Our benchmarks were executed on hardware nodes equipped with a single 64-core AMD EPYC “Trento” CPU and four AMD MI250X GPUs.

\subsection{Cold-start QAS}
\label{sec:cold}

We consider the curriculum reinforcement learning (CRL) framework for quantum architecture search (QAS)~\cite{patel2024curriculum}. Within CRLQAS, the agent interacts with an environment where quantum circuits are encoded as 3D binary tensors, capturing gate operations across qubits, circuit depth, and gate types. The action space consists of $3N + 2\binom{N}{2}$ possible actions, consisting $\{\texttt{CNOT}, \texttt{RX}, \texttt{RY}, \texttt{RZ}\}$ gates, with illegal action rules to prevent redundant operations (e.g., consecutive identical gates). We employ $\epsilon$-greedy exploration ($\epsilon(t) = \max(0.05, 0.99995^t)$) and experience replay to balance exploration and exploitation. The setting employs DDQN algorithm with $\epsilon$-greedy policy with $\epsilon$-decaying factor $0.99995$, from an initial value $1$ until it reaches minimum $0.05$. The replay buffer size was fixed at $20000$, and the target network in the DDQN training process was updated after every $500$ steps. The choice of DDQN stems from the fact that for quantum chemistry simulations it outperforms other RL-algorithms, providing the most compact (in terms of number of gates and depth) quantum circuits with errors below chemical accuracy~\cite{ikhtiarudin2025benchrl}.

The RL-state consists of a $D_\text{max}\times ((N\times N_{1q})\times N)$ elements, where $D_\text{max}$ is the number of steps per episode (also the maximum allowed depth of the PQC).  
The curriculum mechanism dynamically adjusts optimization difficulty using a feedback-driven threshold $\xi$, calculated from a theoretical lower bound $\mu = -\sum_i |c_i|$ (where $c_i$ are the Pauli coefficients of the Hamiltonian) and the best-found energy $\xi_2$. Threshold updates follow $\xi_{\text{new}} = |\mu - \xi_2| + \delta$, with adaptive reset rules to escape local minima.

% \vspace{-10pt}
\subsubsection{Noisy scenario}
\label{sec:noisy}

We benchmark QASER in a realistic noisy scenario. Very few works explores the performance of QAS in the presence of noise. In~\cite{du2022quantum} authors benchmark noise from \texttt{IBM Ourense} and ref.~\cite{patel2024curriculum} considers a noise model similar to \texttt{IBM Mumbai}, which outperforms the net-based approach of ref.~\cite{du2022quantum}. Unfortunately, both of these QPUs are currently deprecated, so we considered a standard 1- and 2-qubit depolarizing noise model.

Figs.~\ref{fig:noisy_simulation} and ~\ref{fig:noisy_simulation_performance} illustrate the comparative performance comparison between CRLQAS with the reward function from Eq.~\ref{eq:linear_reward_increment}, and CRLQAS with the exponential reward from Eq.~\ref{eq:exp_reward_orig} i.e. \texttt{QASER}. Both are trained for $\sim48$ hours. Fig.~\ref{fig:noisy_simulation} presents the results where we aim to prepare the circuit representing the ground state of $\lih$ and $\smallhtwoo$ in the presence of 1-, and 2-qubit depolarizing noise with strengths $10^{-3}$ and $10^{-2}$, respectively. Fig.~\ref{fig:noisy_simulation_performance} (left panel) shows that both methods maintain a consistent error threshold of approximately $10^{-1}$ until episode $2000$ (which defines the exploration phase of the $\epsilon$-greedy policy), at which point they exhibit a significant performance transition. Post-exploration, QASER demonstrates better convergence of the ground state approximation error, achieving a threshold $\sim 8\times10^{-3}$, while CRLQAS stabilizes at $\sim 5\times10^{-2}$. Meanwhile, Fig.~\ref{fig:noisy_simulation_performance} (right panel) demonstrates that QASER provides the RL agent with a reward signal that is approximately $10\times$ stronger.

\subsubsection{Noiseless scenario}
\label{sec:noiseless}

We evaluate QASER against CRLQAS in a noiseless scenario to prepare the ground state of $\lih$, $\smallhtwoo$, and $\bightwoo$. The results are presented in Table~\ref{tab:noiseless_comparison}.

For $\lih$, QASER achieved an average (across 10 random initializations of the neural network) error of $6.5 \times 10^{-5}$, which is an order of magnitude lower than $8.39 \times 10^{-5}$ achieved by CRLQAS. This performance advantage persisted for larger systems. For $\smallhtwoo$, QASER maintained a lower error $4.3 \times 10^{-4}$ compared to $8.77 \times 10^{-4}$ achieved by CRLQAS. For larger $\bightwoo$ molecule, both methods achieved similar error rates, with QASER achieve $3.4 \times 10^{-4}$ and CRLQAS $3.53 \times 10^{-4}$. However, QASER continue to achieve greater resource efficiency, using an average of 103.1 CNOT gates and an average depth of 81.3, while CRLQAS required 112.5 CNOT gates and a depth of 86.6 on average.

To present robustness, we evaluate QASER against non-RL quantum circuit synthesis methods. The results are presented in Table~\ref{tab:nonrl_comparison}. QASER achieved an average error of $5.09 \times 10^{-5}$, which is significantly lower than the errors obtained by hardware-efficient ansatz (HEA, $6.2 \times 10^{-4}$), training-free quantum architecture search (TF-QAS, $1.8 \times 10^{-3}$), parameterized quantum ansatz search (PQAS, $4.8 \times 10^{-3}$), and random search (RS, $5.0 \times 10^{-3}$). In addition to this accuracy improvement, QASER required substantially fewer gates, with an average of only 48.1 gates, compared to 72 for HEA and 57 for the other methods.

\begin{table}[h!]
    \centering
    \small
    \begin{tabular}{@{}lll@{}}
    \toprule
    % \multicolumn{1}{c}{Average} &
    \textbf{Method}    &   \textbf{Av. Error}   &   \textbf{avg. GATE}     \\ \midrule
    \cellcolor{qaserc!80}\textbf{QASER}    &  \cellcolor{qaserc!80}$\mathbf{5.09 \times 10^{-5}}$ & \cellcolor{qaserc!80}\textbf{48.1} \\
    TF-QAS~\cite{he2024training}      &   $1.8 \times 10^{-3}$ & 57\\
    HEA-4~\cite{kandala2017hardware}      &   $6.2 \times 10^{-4}$ & 72\\
    PQAS~\cite{zhang2021neural}      &   $4.8 \times 10^{-3}$ & 57\\
    Random sampling      &   $5.0 \times 10^{-3}$ & 57\\
    \bottomrule
    \end{tabular}

\vspace{5mm}
\caption{Comparison with non-RL approaches: QASER obtains lower error with more compact quantum circuits (at least 20\% in the number of gates) for the $\behtwo$ molecule. HEA-4 represents hardware efficient ansatz with 4 layers.}
\label{tab:nonrl_comparison}
\end{table}
\subsection{Warm-start QAS}
\label{sec:warm}

\begin{figure}[t!]
    \centering
    \includegraphics[width=.75\linewidth]{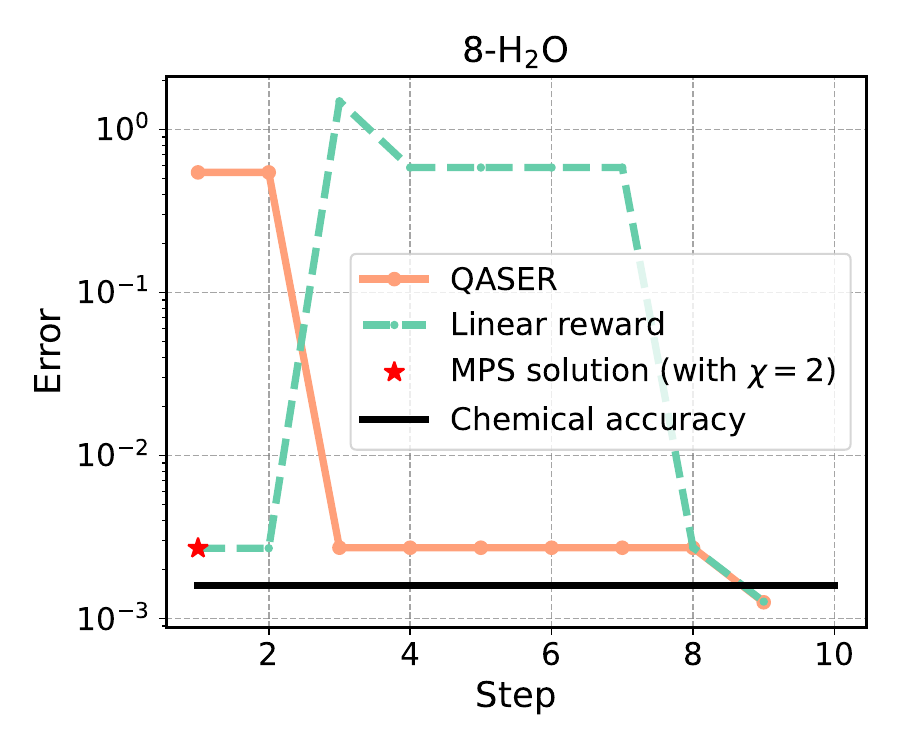}
    \caption{A typical episode in TensorRL with two rewards: QASER (Eq.~\ref{eq:exponential_reward_increment}) and the linear reward (Eq.~\ref{eq:linear_reward_increment}). QASER achieves faster and more stable convergence to low energy estimations, although it is not starting from the MPS state.}
    \label{fig:stability_in TensorRL}
\end{figure}

During the preparation of this research, TensorRL-QAS~\cite{kundu2025tensorrlqas}, a new variant of RL-QAS, was introduced. TensorRL-QAS is based on tensor networks and has been shown to scale up to 20-qubits. The framework leverages tensor network representations such as matrix product states (MPS) to warm-start the RL-agent: having the initial solution already close to the goal, the RL-agent's primary task is to discover circuits which improve the MPS solution, rather than start from scratch. TensorRL-QAS uses a vanilla RL-curriculum and a linear reward function similar to the one presented in Eq.~\ref{case:error_rwd_CRLQAS_vanilla}.

Because the MPS initialization already ensures a compact, efficient representation, and further RL-driven gate additions mostly serve to fine-tune the solution, strict penalties for circuit size or depth can limit the agent's ability to make crucial improvements~\cite{moflic2023cost}. Imposing these constraints risks discouraging any additional exploration. Hence, to leverage the benefits of an exponential reward, we replace the intermediate reward from 
\begin{equation}
    \mathcal{E}_\text{linear} = \max\left( \frac{E_{t-1} - E_t}{|E_{t-1} - E_\text{min}|}, -1 \right),
    \label{eq:linear_reward_increment}
\end{equation}
to an exponential form:
\begin{equation}
\mathcal{E}_\text{QASER} = \exp\left( \alpha \max\left( \frac{E_{t-1} - E_t}{|E_{t-1} - E_\text{min}|},\ -1 \right) \right),
\label{eq:exponential_reward_increment}
\end{equation}
where $E_t$ is the energy at step $t$, $E_{t-1}$ is the previous energy, $E_\text{min}$ is the minimum (target) energy, and $\alpha$ is a scaling hyperparameter. Note that when $E_{t-1} \to 0$, $\mathcal{E}_\text{QASER}$ = $\exp({\alpha \max(\frac{-E_{t-1}}{E_{min}}},-1))$ which is a scaled version of the function from Eq.~\ref{eq:exp_reward_orig} which ignores circuit-related costs.

The $\mathcal{E}_\text{QASER}$ function systematically incentivizes incremental and sustained reductions in energy error at each optimization step, promoting a smooth and stable improvement of error over the MPS solution as gates are sequentially added at each step. Notably, the initial quantum gate application results in an increase in error relative to the MPS solution, but subsequent steps either maintain or further reduce the error, supporting consistent convergence without abrupt error spikes.

In contrast, the linear reward function $\mathcal{E}_\text{linear}$ emphasizes substantial improvements in error, rewarding the agent primarily for large breakthroughs. Minor improvements or plateaus receive little recognition, which can lead to less stable error profiles characterized by sudden drops and occasional increases from riskier actions.

\begin{table}[h!]
\centering
\small
    \begin{tabular}{lcccc}
    \toprule
    \textbf{Model (best)} & \textbf{Error} & \textbf{Depth} & \textbf{CX} & \textbf{ROT} \\
    \midrule
    Min. TensorRL-QAS~\cite{kundu2025tensorrlqas}     & $\mathbf{0.89\times 10^{-3}}$ & 6 & 9 & 15 \\
    Min. TensorRL-QAS (rerun)     & $0.97\times 10^{-3}$ & 5 & \textbf{4} & 5  \\
    Min. \cellcolor{qaserc!80}\textbf{TensorRL-QAS (QASER)}     & $1.0\times 10^{-3}$  & \cellcolor{qaserc!80}\textbf{4} & 9 & \cellcolor{qaserc!80}\textbf{4}  \\
\midrule
    Avg. TensorRL-QAS (rerun) & $1.2\times10^{-3}$ & $5.14$ & $5.43$ & $3.14$ \\
    \cellcolor{qaserc!80}Avg. TensorRL-QAS (QASER)  & $1.2\times10^{-3}$ & $7.14$ & $8.57$ & $6.43$ \\
    \bottomrule
    \end{tabular}
    
\vspace{5mm}
\caption{\textit{TensorRL-QAS (QASER)} is the TensorRL-QAS with our exponential intermediate increment. We compare it with the original results from~\cite{kundu2025tensorrlqas}, and also a rerun presented in \textit{TensorRL-QAS (rerun)}. The best values are marked by Min. and the averages over 10 independent seeds by Avg.}
\label{tab:reward_comparison_for_TensorRL}
\end{table}

To investigate the stability of QASER under TensorRL-QAS, we consider the task of preparing the ground state of $\smallhtwoo$ molecule. We warm start the RL-state using an MPS~\cite{cirac2021matrix} (with bond dimension $\chi=2$) approximation of the ground state. The primary result is presented in Fig.~\ref{fig:stability_in TensorRL} where we show a typical RL-episode during the training. Both QASER (Eq.~\ref{eq:exponential_reward_increment}) and the linear reward (Eq.~\ref{eq:linear_reward_increment}) reach chemical accuracy in the same number of steps. However, QASER consistently reduces errors after the initial step, whereas the linear reward induces substantial error oscillations within the episode. This demonstrates that QASER produces smoother and more reliable learning dynamics, outperforming in terms of robustness and optimization stability.

Furthermore, we train TensorRL-QAS for 10 different initializations of the neural network using both linear and QASER reward schemes. As the performance of same algorithm under same seed might vary under different GPUs, we rerun the TensorRL-QAS highlighted as \texttt{TensorRL-QAS (rerun)}. Single-seed results for the best observed circuits are provided in the top part of Table~\ref{tab:reward_comparison_for_TensorRL}, showing that the original method~\cite{kundu2025tensorrlqas} achieves the lowest error ($0.89\times 10^{-3}$) with deeper, more complex circuits, while the rerun (on current CPU and GPU hardware) with linear reward yields slightly higher error ($0.97\times 10^{-3}$) with improved gate efficiency ($4$ CX, $5$ ROT). QASER enables competitive accuracy ($1.0\times 10^{-3}$) alongside the shallowest and most 1-qubit efficient circuit.

\section{Conclusion}

Quantum architecture search (QAS) aims to automatically construct quantum circuits that achieve high accuracy while remaining shallow enough for realistic, error-prone hardware. Existing RL-based QAS struggle with simultaneous optimization of circuit depth and problem accuracy as their reward functions typically optimize a single metric. We address this by introducing QASER, an exponential, object-oriented reward formulation that jointly accounts for circuit depth, entangling gate cost, and accuracy through a max-tracking mechanism that stabilizes learning and prevents reward hacking. Integrated into cold-start and warm-start QAS frameworks, QASER discovers circuits with up to 20\% fewer 2-qubit gates, reduced depth, and up to 50\% higher accuracy compared to state-of-the-art RL-based methods. On quantum chemistry benchmarks, QASER achieves the same or lower energy error in preparing ground state for chemical Hamiltonian than prior approaches while generating significantly more resource-efficient circuits. Moreover, QASER improves convergence speed in warm-start QAS. These results demonstrate that reward engineering, rather than architectural changes alone, can advance hardware-aware quantum architecture search.

\balance

\bibliographystyle{unsrt}
\bibliography{ref}

\end{document}